\documentstyle[preprint,aps]{revtex}
\begin{document}

\draft

\title{Eigen-functional bosonization and Eikonal-type equations in 
one-dimensional strongly correlated electron systems}

\author{Yu-Liang Liu}
\address{Center for Advanced Study, Tsinghua University, Beijing 100084, 
People's Republic of China}

\maketitle

\begin{abstract}

With the eigen-functional bosonization method, we study one-dimensional
strongly correlated electron systems with large momentum ($2k_{F}$ and/or
$4k_{F}$) transfer term(s), and demonstrate that 
this kind of problems ends in to
solve the Eikonal-type equations, and these equations are universal, and
independent of whether or not the system is integrable. In contrast to
usual perturbation theory, this method is valid not only for weak electron 
interaction, but also for strong electron interaction. Comparing with
exact solution of some integrable models, it can give
correct results in one-loop approximation. This method can also be used to
study electron-phonon interaction systems, and two coupled spin chain or
quantum wire systems. 

\end{abstract}
\vspace{1cm}


\newpage

\section{Introduction}

In contrast to two- and three-dimensional electron systems, one-dimensional
electron system has a prominent feature that its Fermi ``surface'' is 
composed of two points $\pm k_{F}$, defined by the Fermi momentum $k_{F}$,
and its Hilbert space is strongly suppressed.
For a one-dimensional electron gas, 
there only exist two kinds of electron excitations:
one is near these two Fermi levels $\pm k_{F}$, respectively, usually 
called as small momentum excitation, and another one is the excitation
between these two
Fermi levels, called as large (crossover) momentum excitation. 
If there is only the former excitation, the system generally shows the 
Luttinger liquid behavior. After turning on the latter excitation, the system
becomes an insulator for repulsive electron interaction.

In low energy regime, the physics property of a one-dimensional electron
system is determined by the electron states near its two Fermi levels 
$\pm k_{F}$. Therefore, it is reasonable to linearize electron energy
spectrum near these two Fermi levels\cite{4,5,6} with a band-width parameter
$D$, and the electrons are divided into two parts: one presents the electrons
near the Fermi level $+k_{F}$, called right-moving electrons with the energy
spectrum $\epsilon_{R}(k)=v_{F}k$, and another
one presents the electrons near the Fermi level $-k_{F}$, called left-moving
electrons with the energy spectrum $\epsilon_{L}(k)=-v_{F}k$, where $v_{F}$
is the electron Fermi velocity. It is shown\cite{6} that the high order terms
like $k^{n}$, $n=$2,3,..., only contribute high order irrelevant terms to the
effective action of the system, and can be neglected. 
The crossover (large momentum) excitation
corresponds to the excitation between the rigth- and left-moving electrons
with $2k_{F}$ and/or $4k_{F}$ (or higher) momentum transfer. 
In some special cases, the crossover excitation is very important, and 
determines the low energy behavior of the system, because the crossover
excitation term is relevant, and strongly affected by the electron interaction.
Generally, usual perturbation methods fail to treating the crossover
excitation term for the strong electron correlation.

The bosonization method is very powerful in studying of 
one-dimensional interacting
electron systems, such as quantum wires, spin chains and impurity scattering
in one-dimensional fermion systems, where it can exactly
treat electron's density-density interactions\cite{4,5,6,7,8}, such as  
the interaction $V(x-y)\rho_{R\sigma}(x)\rho_{L\sigma}(y)$.
However, for example, if there meantime appear
an interaction term $V\rho_{R\sigma}(x)\rho_{L\sigma}(x)$ and a crossover term
$\lambda[\psi^{\dagger}_{R\sigma}(x)\psi_{L\sigma}(x)+
\psi^{\dagger}_{L\sigma}(x)\psi_{R\sigma}(x)]$
in the Hamiltonian of the system, it is hard to use
the usual bosonization and functional bosonization methods to
effectively treat these two terms as a whole, because the $\lambda$-term mixes
the right- and left-moving electrons which induces a highly relevant term,
in the bosonization representation of the electron fields, it reduces to some
cosine- and/or sine- terms. The usual bosonization methods can only tell us
which ones are relevant or irrelevant, but cannot be used to effectively 
treat these terms, such as calculating spin and charge collective excitation
spectrums, correlation functions, effective action of the system, and so on.
Therefore, the usual bosonization methods
do not provide more useful informations than usual perturbation methods.
It is a big challenge to effectively treat these crossover excitation terms
which often appear in one-dimensional strongly correlated electron systems,
and desires to find a new method which can be used to exactly and effectively
treat them.

In this paper, we introduce a new method\cite{8'},
called the eigen-functional bosonization method, which is very simple and 
effective to treating these crossover terms, and show that 
the crossover terms can be exactly treated by solving the Eikonal-type
equations. It is wonderful that the problems of one-dimensional strongly
correlated electron systems end in to solve the Eikonal-type equations that
can be exactly treated by a series expansion and/or computer calculations.
This method is universal, and independent of whether or not the 
system is integrable. For example, it can also be used to treat 
electron-phonon interaction systems, and two- and three-dimensional electron
systems\cite{8''}. In Sect. II, for simplicity,
we introduce the eigen-functional bosonization method by studying
a one-dimensional spinless fermion system with a staggered chemical
potential or a dimerized lattice sites, and show that this kind of problem
ends in to solve simple Eikonal-type equations. At one-loop approximation,
our calculation is consistent with the exact solution of the quantum
sine-Gordon model. In Sect. III, with
this method, we study the Hubbard model at half filling, and calculate
charge collective excitation gap which is agreement with the exact solution
of the Hubbard model at half filling. In Sect. IV, we study the Hubbard
model at half filling with a staggered chemical potential, and show that
there is a phase transition from band insulator to Mott-type 
insulator at the electron interaction strength $V_{T}$ where the charge
collective excitation gap is zero, and a spontaneous lattice dimerization
takes place. This property of the system can be used to explain the origin 
of the ferroelectricity of transition metal oxides\cite{9'}.
The conclusion and discussions are given in Sect. V. 

\section{eigen-functional bosonization and its application}

In this section, we introduce the eigen-functional bosonization method by
studying a one-dimensional spinless fermion system with a staggered chemical
potential or a dimerized lattice sites. It is well-known that in usual 
bosonization representation of fermion fields, the system reduces into the 
quantum sine-Gordon model (QSGM) which is integrable, and has been exactly 
solved\cite{1}. The exact solution of the QSGM is a criterion for this 
new method. We show that in one-loop approximation this method can give
correct results for any fermion interaction strength, whereas usual 
perturbation methods are invalid for strong fermion interaction.
 
Generally, we consider a strongly correlated spinless fermion system with the 
Hamiltonian
\begin{eqnarray}
H &=& \displaystyle{-i\hbar v_{F}\int dx [\psi^{\dagger}_{R}(x)\partial_{x}
\psi_{R}(x)-\psi^{\dagger}_{L}(x)\partial_{x}\psi_{L}(x)]} \nonumber \\
&-& \displaystyle{ \int dx [\lambda\psi^{\dagger}_{R}(x)\psi_{L}(x)+
\lambda^{*}\psi^{\dagger}_{L}(x)\psi_{R}(x)] 
+ V\int dx \rho_{R}(x)\rho_{L}(x)}
\label{1}\end{eqnarray}
where $\psi_{R(L)}(x)$ is the right(left)-moving fermion field, and
$\rho_{R(L)}(x)=\psi^{\dagger}_{R(L)}(x)\psi_{R(L)}(x)$ is the 
right(left)-moving fermion density. The $\lambda$-term represents a staggered
chemical potential ($\lambda=\lambda^{*}=\Delta_{0}$), 
or a lattice dimerization parameter ($\lambda=-\lambda^{*}=iu$).
Generally, we can write $\lambda$ as $\lambda=\Delta_{0}+iu$.
We can also study the influence of quantum fluctuation of 
electron-acoustic-phonon interaction on the lattice dimerization by adding 
the term, $u_{a}(x)[\rho_{R}(x)+\rho_{L}(x)]$.
In bosonization representation,
$\psi_{R(L)}(x)=(\frac{1}{2\pi\alpha})^{1/2}exp\{-i\Phi_{R(L)}(x)\}$, where
$\partial_{x}\Phi_{R(L)}(x)=\pm 2\pi\rho_{R(L)}(x)$, $\alpha\sim 1/D$, and
$D$ is the band-width, we have the standard QSGM\cite{2} ($u=0$),
\begin{eqnarray} 
H_{sG}=\frac{\hbar v_{c}}{2}\int dx\{[\partial_{x}\Phi(x)]^{2}+
\Pi^{2}(x)+\frac{m^{2}}{\beta^{2}}\cos(\beta\Phi(x))\} \nonumber
\end{eqnarray}
where, $m^{2}/\beta^{2}=\lambda/(\pi\alpha\hbar v_{c})$, $\beta^{2}/(4\pi)=
(1-\gamma)^{1/2}/(1+\gamma)^{1/2}=g$, $v_{c}=v_{F}\sqrt{1-\gamma^{2}}$,
$\gamma=V/(2\pi\hbar v_{F})$, $\Phi(x)=[\Phi_{R}(x)-\Phi_{L}(x)]/\beta$,
$\Pi(x)=\beta[\partial_{x}\Phi_{R}(x)+\partial_{x}\Phi_{L}(x)]/(4\pi)$, and
$[\Pi(x), \Phi(y)]=i\delta(x-y)$. The parameter $m$ is related to the mass
of the solitons and anti-solitons or breathers\cite{2}.
According to the relation $\beta^{2}=4\pi g$, at
$\beta^{2}=4\pi$, the Hamiltonian (\ref{1}) describes a free massive fermion
system ($V=0$). As $\beta^{2}>4\pi$ ($g>1$), there is an attractive 
interaction between the fermions $\psi_{R}(x)$ and $\psi_{L}(x)$. However, 
due to the
crossover term ($\lambda$-term) opens a gap in fermion excitation
spectrum, the 
system is stable. As $\beta^{2}<4\pi$ ($g<1$), the fermion interaction is 
repulsive. In the fermionization representation\cite{9,10}, 
it is very clear that 
in the regime $0<\beta^{2}<4\pi$ the
QSGM represents different physics from that in the regime 
$4\pi<\beta^{2}<8\pi$.
The critical point $\beta^{2}=4\pi$ is a phase transition point 
which separates one phase from another one.

After introducing auxiliary boson fields $\phi_{R}(x,t)$ and $\phi_{L}(x,t)$
which decouple the four-fermion interaction by adding the constraints,
$\rho_{R(L)}(x,t)=\psi^{\dagger}_{R(L)}(x,t)\psi_{R(L)}(x,t)$, we only have
the quadratic terms of the fermion fields $\psi_{R(L)}(x,t)$. 
Integrating out them, we obtain an effective action,
\begin{eqnarray}
S[\rho,\phi] &=& \displaystyle{ 
W[\phi]-\int dt dx \{ \phi_{R}(x,t)\rho_{R}(x,t)+\phi_{L}(x,t)
\rho_{L}(x,t)+V\rho_{R}(x,t)\rho_{L}(x,t)\}} \nonumber \\
W[\phi] &=& \displaystyle{ i Tr\ln(M_{0})+i\int^{1}_{0}d\xi\int dt dx\;
tr\left(\phi(x,t)\tilde{G}(x,t;x',t',[\xi\phi])\right)|_{\stackrel{t'
\rightarrow t}{x'\rightarrow x}}}
\label{2}\end{eqnarray}
where $M=M_{0}+\phi$, ${\cal D}_{R(L)}=i\partial_{t}\pm i\hbar v_{F}
\partial_{x}$, $M\tilde{G}(x,t;x',t',[\phi])=\delta(x-x')\delta(t-t')$, and
\begin{eqnarray}
M_{0}=\left( \begin{array}{c} \displaystyle{ {\cal D}_{R}},  
\; \lambda \\ \lambda^{*},  \; \displaystyle{{\cal D}_{L}}
\end{array}\right), \;\;\;\;
\phi(x,t)=\left( \begin{array}{ll} \displaystyle{ \phi_{R}(x,t)}, 
\; 0\\ 0,  \; \displaystyle{\phi_{L}(x,t)} \end{array}\right).
\nonumber\end{eqnarray}
In order to calculate the Green's functional $\tilde{G}(x,t;x',t',[\phi])$, 
we solve the eigen-functional equation\cite{10'} of the operator $M$,
\begin{equation}
M\Psi^{(i)}_{k,\omega}(x,t,[\phi])=E^{(i)}_{k\omega}[\phi]
\Psi^{(i)}_{k,\omega}(x,t,[\phi])
\label{3}\end{equation}
With the orthogonality and normalization of the eigen-functionals,
\begin{eqnarray}
\displaystyle{\sum_{k,\omega}\Psi^{(i)\dagger}_{k,\omega}(x,t,[\phi])
\Psi^{(i')}_{k,\omega}(x',t',[\phi])} &=& \delta_{ii'}\delta(x-x')
\delta(t-t'), \nonumber \\
\displaystyle{\int dt dx \Psi^{(i)\dagger}_{k,\omega}(x,t,[\phi])
\Psi^{(i')}_{k',\omega'}(x,t,[\phi])} &=& \delta_{ii'}\delta_{kk'}
\delta_{\omega\omega'}
\nonumber\end{eqnarray}
the Green's functional can be written as,
\begin{equation}
\tilde{G}(x,t;x',t',[\phi])=\sum_{i}\sum_{k,\omega}\frac{1}{
E^{(i)}_{k\omega}[\phi]}
\Psi^{(i)}_{k,\omega}(x,t,[\phi])\Psi^{(i)\dagger}_{k,\omega}
(x',t',[\phi])
\label{4}\end{equation}
If we can exactly solve the eigen-functional equation (\ref{3}), generally
we can calculate the Green's functional (\ref{4}), and the effective action
(\ref{2}) which only includes the boson fields. This process is called the
eigen-functional bosonization.

At $\phi(x,t)=0$ (interacting-free case), the system becomes a band insulator,
and the eigen-equation (\ref{3}) can be exactly solved. It 
has two sets of 
solutions corresponding to the eigen-values $E^{(\pm)}_{k\omega}[0]
=\omega\mp E_{k}$,
respectively, where $E_{k}=\sqrt{k^{2}+|\lambda|^{2}}$ (choosing $\hbar=v_{F}
=1$), 
\begin{eqnarray}
\Psi^{(+)}_{k,\omega}(x,t,[0]) &=& \displaystyle{ (\frac{1}{TL})^{1/2}
\left( \begin{array}{c} u_{k}e^{i\theta}
\\ -v_{k}e^{-i\theta}\end{array}\right)e^{ikx-i\omega t}}
\nonumber \\
\Psi^{(-)}_{k,\omega}(x,t,[0]) &=& \displaystyle{ (\frac{1}{TL})^{1/2}
\left( \begin{array}{c} v_{k}e^{i\theta}
\\ u_{k}e^{-i\theta}\end{array}\right)e^{ikx-i\omega t}}
\label{4a}\end{eqnarray}
where $u^{2}_{k}=1-v^{2}_{k}=(1+k/E_{k})/2$, $\tan(2\theta)=u/\Delta_{0}$,
and $T$ and $L$ are the time and
space lengths of the system, respectively. 

At $\lambda=0$, the Hamiltonian (\ref{1}) represents the Luttinger liquid, and
the eigen-functional equation (\ref{3}) has the exact solutions with the
eigen-values $E^{(\pm)}_{k\omega}=\omega\mp |k|$, respectively,
\begin{equation}
\Psi^{'(\pm)}_{k,\omega}(x,t,[\phi])=\left(\frac{1}{TL}\right)^{1/2}
\left( \begin{array}{c}
\displaystyle{\theta(\pm k)e^{Q^{0}_{R}(x,t)}}\\
\displaystyle{\mp\theta(\mp k)e^{Q^{0}_{L}(x,t)}}\end{array}\right)
e^{ikx-i\omega t}
\label{4b}\end{equation}
where ${\cal D}_{R(L)}Q^{0}_{R(L)}(x,t)=-\phi_{R(L)}(x,t)$, ant $\theta(x)$
is a step function, $\theta(x)=1$ for $x>0$, and $\theta(x)=0$ for $x<0$.
With the exact solution (\ref{4b}), we can easily obtain the effective action
of the system, and the fermion Green's function which are the same as that
in Ref.\cite{7,8}.

In general, for $\phi(x,t)\neq 0$, we can choose 
the following formal solutions with the 
eigen-values $E^{(\pm)}_{k\omega}[\phi]
=\omega\mp E_{k}+\Sigma^{(\pm)}_{k}[\phi]$, respectively,
\begin{eqnarray}
\Psi^{(+)}_{k,\omega}(x,t,[\phi]) &=& \displaystyle{ A_{k}(\frac{1}{TL})^{1/2}
\left(\begin{array}{c} \displaystyle{u_{k}e^{i\theta}e^{Q_{R}(x,t)}}\\
\displaystyle{-v_{k}e^{-i\theta}
e^{Q_{L}(x,t)}}\end{array}\right)e^{ikx-i(\omega+
\Sigma^{(+)}_{k}[\phi])t}}, \nonumber \\
\Psi^{(-)}_{k,\omega}(x,t,[\phi]) &=& \displaystyle{ A_{k}(\frac{1}{TL})^{1/2}
\left(\begin{array}{c} \displaystyle{v_{k}e^{i\theta}e^{\tilde{Q}_{R}(x,t)}}\\
\displaystyle{u_{k}e^{-i\theta}e^{\tilde{Q}_{L}(x,t)}}\end{array}\right)
e^{ikx-i(\omega+\Sigma^{(-)}_{k}[\phi])t}},
\label{5}\end{eqnarray}
where $\Sigma^{(\pm)}_{k}[\phi]=
\int^{1}_{0}d\xi\int dt dx\Psi^{(\pm)\dagger}_{k,\omega}(x,t,[\xi\phi])
\phi(x,t)\Psi^{(\pm)}_{k,\omega}(x,t,[\xi\phi])$ are independent of $\omega$,
and $|A_{k}|\sim 1$ is a normalization constant. 
The boson fields $Q_{R(L)}(x,t)$ and $\tilde{Q}_{R(L)}(x,t)$ satisfy the 
differential equations,
\begin{eqnarray}
&&\left\{ \begin{array}{ll}
\displaystyle{ ({\cal D}_{R}+\phi_{R}(x,t))e^{Q_{R}(x,t)}-\frac{|\lambda|
v_{k}}{u_{k}}(e^{Q_{L}(x,t)}-e^{Q_{R}(x,t)}) } &= 0,\\
\displaystyle{ ({\cal D}_{L}+\phi_{L}(x,t))e^{Q_{L}(x,t)}-\frac{|\lambda|
u_{k}}{v_{k}}(e^{Q_{R}(x,t)}-e^{Q_{L}(x,t)})} &= 0,\end{array}\right.
\nonumber \\
&&\left\{ \begin{array}{ll}
\displaystyle{ ({\cal D}_{R}+\phi_{R}(x,t))e^{\tilde{Q}_{R}(x,t)}+
\frac{|\lambda| u_{k}}{v_{k}}(e^{\tilde{Q}_{L}(x,t)}-e^{\tilde{Q}_{R}(x,t)})}
= & 0, \\
\displaystyle{ ({\cal D}_{L}+\phi_{L}(x,t))e^{\tilde{Q}_{L}(x,t)}+
\frac{|\lambda| v_{k}}{u_{k}}(e^{\tilde{Q}_{R}(x,t)}-e^{\tilde{Q}_{L}(x,t)})}
= & 0, \end{array}\right.
\label{6}\end{eqnarray}
With the following definitions
\begin{eqnarray}
Q_{R(L)}(x,t)=Q^{0}_{R(L)}(x,t)+f_{R(L)}(x,t), \;\;\;\;
\tilde{Q}_{R(L)}(x,t)=Q^{0}_{R(L)}(x,t)+\tilde{f}_{R(L)}(x,t)
\nonumber\end{eqnarray}
the differential equation (\ref{6}) can be re-written as,
\begin{eqnarray}
\left\{ \begin{array}{ll}
\displaystyle{ \left( {\cal D}_{L}{\cal D}_{R}+iE_{k}\partial_{t}+ik
\partial_{x}\right)f_{R}+{\cal D}_{R}f_{R}{\cal D}_{L}f_{R}} = &
\displaystyle{ -\left(\frac{|\lambda| v_{k}}{u_{k}}+{\cal D}_{R}f_{R}\right)
{\cal D}_{L}Q^{0}}\\
\displaystyle{ \left( {\cal D}_{R}{\cal D}_{L}+iE_{k}\partial_{t}+ik
\partial_{x}\right)f_{L}+{\cal D}_{R}f_{L}{\cal D}_{L}f_{L}} = &
\displaystyle{  \left(\frac{|\lambda| u_{k}}{v_{k}}+{\cal D}_{L}f_{L}\right)
{\cal D}_{R}Q^{0}} \end{array}\right. \nonumber \\
\left\{ \begin{array}{ll}
\displaystyle{ \left( {\cal D}_{R}{\cal D}_{L}-iE_{k}\partial_{t}+ik
\partial_{x}\right)\tilde{f}_{R}+{\cal D}_{R}\tilde{f}_{R}{\cal D}_{L}
\tilde{f}_{R}} = & \displaystyle{ \left( \frac{|\lambda| u_{k}}{v_{k}}-
{\cal D}_{R}\tilde{f}_{R}\right){\cal D}_{L}Q^{0}}\\
\displaystyle{ \left( {\cal D}_{R}{\cal D}_{L}-iE_{k}\partial_{t}+ik
\partial_{x}\right)\tilde{f}_{L}+{\cal D}_{R}\tilde{f}_{L}{\cal D}_{L}
\tilde{f}_{L}} = & \displaystyle{-\left( \frac{|\lambda| v_{k}}{u_{k}}-
{\cal D}_{L}\tilde{f}_{L}\right){\cal D}_{R}Q^{0}} \end{array}\right.
\label{7}\end{eqnarray}
where $Q^{0}(x,t)=Q^{0}_{R}(x,t)-Q^{0}_{L}(x,t)$. This is usual Eikonal-type 
equations\cite{fradkin,11}, 
and can be solved by a series expansion of $Q^{0}(x,t)$, or by
computer calculations.
These equations are universal, and independent of whether or not the system
is integrable. For example, after including the electron-acoustic-phonon
interaction, they keep invariance if one replaces 
$\phi_{R(L)}(x,t)$ by $\phi_{R(L)}(x,t)-u_{a}(x,t)$. 
 
The non-linear terms will produce the high-order terms of $(|\lambda|
/\hbar v_{F})Q^{0}(x,t)$, and contribute cubic or higher order terms of
the boson fields $\phi_{R(L)}(x,t)$ to the effective action,
as $|\lambda|/\hbar v_{F}\ll 1$ they can be neglected.
Only keeping the linear terms (one-loop approximation), we can obtain 
\begin{eqnarray}
\displaystyle{ \left( \begin{array}{c}
f_{R}(q,\Omega)\\ f_{L}(q,\Omega)\end{array}\right)} &=& \displaystyle{
\frac{1}{\Omega^{2}-q^{2}+2(E_{k}\Omega-kq)}\left( \begin{array}{c}
\frac{|\lambda| v_{k}}{u_{k}}(\Omega+q)\\ -\frac{|\lambda| u_{k}}{v_{k}}
(\Omega-q)\end{array}\right)Q(q,\Omega)} \nonumber \\
\displaystyle{ \left( \begin{array}{c}
\tilde{f}_{R}(q,\Omega)\\ \tilde{f}_{L}(q,\Omega)\end{array}\right)} &=&
\displaystyle{ \frac{1}{\Omega^{2}-q^{2}-2(E_{k}\Omega+kq)}\left(
\begin{array}{c} -\frac{|\lambda| u_{k}}{v_{k}}(\Omega+q)\\
\frac{|\lambda| v_{k}}{u_{k}}(\Omega-q)\end{array}\right)Q(q,\Omega)}
\label{8}\end{eqnarray}
where $Q(q,\Omega)=\phi_{R}(q,\Omega)/(\Omega-q)-\phi_{L}(q,\Omega)
/(\Omega+q)$. 
Generally, these boson fields depend on the fermion 
momentum $k$. However, the low energy physical property is determined by
the fermions near the Fermi levels $\pm k_{F}$, we therefore can take
the approximation in (\ref{8}), $k\sim 0$ and $E_{k}\sim |\lambda|$.
Under this approximation, these boson fields are independent of the fermion
momentum $k$. 

According to equations (\ref{4}) and (\ref{5}), the Green's functional
can be written as,
\begin{equation}
\tilde{G}(x,t;x',t,[\phi])=-i \left( \begin{array}{ll}
\displaystyle{G_{R}(x-x')e^{Q_{R}(x,t)-Q_{R}(x',t)}}, \;\; &
\displaystyle{Z(x-x')e^{Q_{R}(x,t)-Q_{L}(x',t)}}\\
\displaystyle{Z(x-x')e^{Q_{L}(x,t)-Q_{R}(x',t)}}, \;\; &
\displaystyle{G_{L}(x-x')e^{Q_{L}(x,t)-Q_{L}(x',t)}}\end{array}\right)
\label{8a}\end{equation}
where $G_{R(L)}(x)=G^{0}_{R(L)}(x)\pm g(x)$, $G^{0}_{R(L)}(x)=\pm 1/(2\pi ix)$,
$g(x)=\frac{1}{4\pi}\int^{0}_{-D}dk(\frac{k}{|k|}-\frac{k}{E_{k}})e^{ikx}$,
and $Z(x)=\frac{1}{4\pi}\int^{0}_{-D}dk\frac{|\lambda|}{E_{k}}e^{ikx}$,
where $D$ is the band-width. Here we have used $Q^{*}_{R(L)}(x,t)
=-Q_{R(L)}(x,t)$, at one-loop approximation this relation is exact.
With the relation\cite{7},
\begin{equation}
\tilde{G}(x,t;x',t',[\phi])|_{\stackrel{t'\rightarrow t}{x'\rightarrow x}}
=\frac{1}{2}\lim_{\eta\rightarrow 0}[\tilde{G}(x,t;x+\eta,t,[\phi])+
\tilde{G}(x,t;x-\eta,t,[\phi])]
\label{8b}\end{equation}
we can obtain the following effective
action (keeping the same zero energy level as fermions, we replace
$\Omega +|\lambda|$ by $\Omega$),
\begin{eqnarray}
S[\phi,\rho] &=& \displaystyle{ \frac{1}{TL}\sum_{q,\Omega}\left\{
-\frac{q}{4\pi}\frac{\Omega+q}{{\cal D}}|\phi_{R}(q,\Omega)|^{2}+
\frac{q}{4\pi}\frac{\Omega-q}{{\cal D}}
|\phi_{L}(q,\Omega)|^{2}\right.} \nonumber \\
&-& \displaystyle{ \left.\phi_{R}(-q,-\Omega)\rho_{R}(q,\Omega)-
\phi_{L}(-q,-\Omega)\rho_{L}(q,\Omega)-V\rho_{R}(-q,-\Omega)
\rho_{L}(q,\Omega)\right\}}
\label{9}\end{eqnarray}
where ${\cal D}=\Omega^{2}-q^{2}-|\lambda|^{2}$. 
With the functional expressions (\ref{5}) of the fermion fields and the 
effective action (\ref{9}), we can calculate any correlation functions of the
system by taking functional average, for example, 
the fermion Green's function is,
\begin{equation}
\tilde{G}(x,t;x',t')= \frac{1}{Z}\int \prod_{i=R,L}D\phi_{i}D\rho_{i}
\tilde{G}(x,t;x',t',[\phi])e^{-iS[\phi,\rho]}
\label{10}\end{equation}
where $Z=\int \prod_{i=R,L}D\phi_{i}D\rho_{i}e^{-iS[\phi,\rho]}$. As $\lambda
=0$, we can obtain the usual fermion Green's function\cite{4,5,6,7,8}. 
At $V=0$, we can obtain the expectation, $<e^{ia(\Phi_{R}(x,t)-\Phi_{L}(x,t))}>
=(\frac{|\lambda|}{D+\sqrt{D^{2}+|\lambda|^{2}}})^{a^{2}}$, 
and for a finite $V$,
$<e^{ia(\Phi_{R}(x,t)-\Phi_{L}(x,t))}>\sim (\frac{|\lambda|}{\alpha D+\sqrt{
\alpha^{2}D^{2}+|\lambda|^{2}}})^{ga^{2}}$, $\alpha^{2}=1-
(\frac{V}{2\pi \hbar v_{F}})^{2}$, which are consistent with the exact
results\cite{ref} where $\varphi=\sqrt{2/g}(\Phi_{R}-\Phi_{L})$.

After integrating out the
auxiliary boson fields $\phi_{R(L)}(x,t)$, we can obtain the collective
excitation spectrums of the fermion density ($\lambda\neq 0$) from the
effective action,
\begin{eqnarray}
\Omega_{1} &=& \displaystyle{ \sqrt{
(1-\frac{1}{2}(\frac{V}{2\pi})^{2})q^{2}+|\lambda|^{2}+F(q,\lambda)}} 
\nonumber \\
\Omega_{2} &=& \displaystyle{ \sqrt{
(1-\frac{1}{2}(\frac{V}{2\pi})^{2})q^{2}+|\lambda|^{2}-F(q,\lambda)}}
\label{11}\end{eqnarray}
where $F(q,\lambda)=\frac{1}{2}|\frac{V}{2\pi}q|\sqrt{(\frac{V}{2\pi})^{2}
q^{2}-4|\lambda|^{2}}$. As $V=0$, the energy
spectrums reduces to $\Omega_{1}=\Omega_{2}=\sqrt{q^{2}+
|\lambda|^{2}}$; as $\lambda=0$, we only have one-branch energy spectrum
$\Omega_{2}=\sqrt{1-(\frac{V}{2\pi})^{2}}|q|$, which is well-knowen in the
usual bosonization representation. Generally, for a finite $\lambda$, the
function $F(q,\lambda)$ becomes imaginary for small $q$ ($|\frac{V}{2\pi}|<1$),
therefore, the small $q$ excitation modes are forbiden. 
The physics meaning of this phonomenon is unclear, it
maybe derives from the finite-size and weak-localization 
of the (anti-) solitons and breathers.

As $\lambda=iu$ and including the static tension potential energy of the
lattices, the dimerization parameter $u$ can be determined by taking
the minimum of the ground state energy of the system\cite{12'}.
When $\lambda=\Delta_{0}$ and taking $\Delta_{0}$ as a free parameter, we can
obtain its renormalization group equation by using the effective 
action (\ref{9}), and study its low energy flow 
behavior in strong repulsive and attractive interaction regions\cite{8'}. 
On the
other hand, the renormalized parameter $\Delta_{0R}$ is related to the mass of 
the (anti-)solitons and breathers in the exact solution of the QSGM.
It is clearly shown\cite{8'} that the fermion interaction dependence of the 
renormalized parameter $\Delta_{0R}$ is completely consistent with the exact
solution of the QSGM. Therefore, we demonstrate that the eigen-functional
bosonization method is powerful, and the low energy behavior of the strongly
correlated systems is controlled by the Eikonal-type equations.
Even if at one-loop approximation it can give
the correct results in treating the one-dimensional strongly correlated 
systems, where in general usual perturbation theory is invalid. 

\section{charge collective excitation gap of the Hubbard model at half-filling}

For the Hubbard model at half filling,
it is well-known\cite{12} that at half-filling, the spin collective
excitation gap is zero, and the charge collective excitation gap is not
zero, and increases with the on-site repulsive Coulomb interaction
potential $U$ of electrons.  
In this section, we use the eigen-functional bosonization method to calculate
the effective action of the electron density fields, so that we can 
determine the electron interaction dependence of the charge collective 
excitation gap of the Hubbard model at half-filling. This system 
represented by the Hubbard model with electron half-filling has a
metal-Mott insulator transition at $U=0$. As $U>0$ it is a Mott insulator,
and at $U=0$ it is a metal. 

In low energy region, and with the linearization of electron spectrum
near the Fermi levels $\pm k_{F}$,
the Hamiltonian of the Hubbard model at half-filling can be written as in the 
continuum limit,
\begin{eqnarray}
H &=& \displaystyle{ -i\hbar v_{F}\sum_{\sigma}\int dx \left[
\psi^{\dagger}_{R\sigma}(x)\partial_{x}\psi_{R\sigma}(x)-
\psi^{\dagger}_{L\sigma}(x)\partial_{x}\psi_{L\sigma}(x)\right]} \nonumber \\
&+& \displaystyle{ V\int dx\left[\rho_{R\uparrow}(x)+\rho_{L\uparrow}(x)
\right]\left[\rho_{R\downarrow}(x)+\rho_{L\downarrow}(x)\right]} \label{23} \\
&+& \displaystyle{\lambda\int dx\left[\psi^{\dagger}_{R\uparrow}(x)
\psi_{L\uparrow}(x)\psi^{\dagger}_{R\downarrow}(x)\psi_{L\downarrow}(x)+
\psi^{\dagger}_{L\uparrow}(x)
\psi_{R\uparrow}(x)\psi^{\dagger}_{L\downarrow}(x)\psi_{R\downarrow}(x)\right]}
\nonumber\end{eqnarray}
where $\psi_{R\sigma}(x)$ ($\psi_{L\sigma}(x)$) is the right (left) moving
electron field with spin $\sigma$, 
$\rho_{R(L)\sigma}(x)=\psi^{\dagger}_{R(L)\sigma}(x)
\psi_{R(L)\sigma}(x)$ is the electron density field, $V=aU$, and $\lambda
=aU$, where $a$ is the lattice constant. Here we have used $\lambda$ to
present the coefficient of the Umklapp scattering which is relevant for 
repulsive electron Coulomb interaction ($U>0$), therefore, in the low
energy region renormalized $\lambda$ is different from $aU$, and neglected the
backward scattering term which is irrelevant for the repulsive interaction
$V>0$. 

With auxiliary fields $\phi_{R\sigma}(x,t)$ and $\phi_{L\sigma}(x,t)$
which introduce the constraints 
$\rho_{R\sigma}(x,t)=\psi^{\dagger}_{R\sigma}(x,t)\psi_{R\sigma}(x,t)$ and 
$\rho_{L\sigma}(x,t)=\psi^{\dagger}_{L\sigma}(x,t)
\psi_{L\sigma}(x,t)$, respectively, and the Hubbard-Stratonovich field
$\Delta_{-\sigma}(x,t)=\psi^{\dagger}_{R\sigma}(x,t)\psi_{L\sigma}(x,t)$,
the action of the system can be written as,
\begin{eqnarray}
S &=& \displaystyle{ \sum_{\sigma}\int dt dx\left\{\Psi^{\dagger}_{\sigma}(x,t)
\hat{M}_{\sigma}(x,t)\Psi_{\sigma}(x,t)\right.} \nonumber \\
&-& \displaystyle{ \left.\phi_{R\sigma}(x,t)\rho_{R\sigma}(x,t)
-\phi_{L\sigma}(x,t)\rho_{L\sigma}(x,t)\right\}} \label{25} \\
&+& \displaystyle{ \int dt dx\left\{\frac{1}{\lambda}\left[
\Delta_{\uparrow}(x,t)\Delta_{\downarrow}(x,t)+\Delta^{*}_{\uparrow}(x,t)
\Delta^{*}_{\downarrow}(x,t)\right]\right.} \nonumber \\
&-& \displaystyle{ \left.V\left[\rho_{R\uparrow}(x,t)+
\rho_{L\uparrow}(x,t)\right]\left[\rho_{R\downarrow}(x,t)+
\rho_{L\downarrow}(x,t)\right]\right\}}
\nonumber\end{eqnarray}
where $\hat{M}_{\sigma}(x,t)=\hat{M}_{0}+\phi_{\sigma}(x,t)$,
$\Psi^{\dagger}_{\sigma}(x,t)=\left(\psi^{\dagger}_{R\sigma}(x,t),
\psi^{\dagger}_{L\sigma}(x,t)\right)$, and 
\begin{eqnarray}
\hat{M}_{0}=\left( \begin{array}{cc}
{\cal D}_{R}, \;\; \Delta\\
\Delta^{*}, \;\; {\cal D}_{L} \end{array}\right), \;\;\;\;
\phi_{\sigma}(x,t)=\left( \begin{array}{cc}
\phi_{R\sigma}(x,t), \;\; \Delta_{\sigma}(x,t)-\Delta\\
\Delta^{*}_{\sigma}(x,t)-\Delta^{*}, \;\; \phi_{L\sigma}(x,t)\end{array}
\right)\nonumber\end{eqnarray}
where ${\cal D}_{R(L)}=i\partial_{t}\pm i\partial_{x}$ (choosing $\hbar=
v_{F}=1$), and $\Delta$ is a parameter (see below). Integrating out the
electron field $\Psi_{\sigma}(x,t)$, we obtain an effective potential,
$W[\phi]=i Tr\ln(\hat{M}_{\sigma})$ which can be represented by using
eigen-functionals,
\begin{eqnarray}
W[\phi] &=& \displaystyle{ i\sum_{\sigma}
\int^{1}_{0} d\xi\int dt dx \; tr\left[\phi_{\sigma}G_{\sigma}(x,t;x',t',
[\xi\phi])\right]|_{\stackrel{t'\rightarrow t}{x'\rightarrow x}}} \nonumber \\
G_{\sigma}(x,t;x',t',[\phi]) &=& \displaystyle{ \sum_{\nu}
\sum_{k,\omega}\frac{1}{E^{(\nu)}_{\sigma k\omega}[\phi]}
\Psi^{(\nu)}_{\sigma k\omega}(x,t,[\phi])\Psi^{(\nu)\dagger}_{\sigma k\omega}
(x',t',[\phi])}
\label{26}\end{eqnarray}
where the eigen-functional $\Psi^{(\nu)}_{\sigma k\omega}(x,t,[\phi])$ 
satisfies the eigen-equation,
\begin{equation}
\hat{M}_{\sigma}(x,t)\Psi^{(\nu)}_{\sigma k\omega}(x,t,[\phi])
=E^{(\nu)}_{\sigma k\omega}[\phi]\Psi^{(\nu)}_{\sigma k\omega}(x,t,[\phi])
\label{27}\end{equation}
Using Hellmann-Feynman theorem, the eigen-value $E^{(\nu)}_{\sigma k\omega}
[\phi]$ reads $E^{(\nu)}_{\sigma k\omega}[\phi]=E^{(\nu)}_{k\omega}[0]+
\Sigma^{(\nu)}_{\sigma k}[\phi]$, where $\Sigma^{(\nu)}_{\sigma k}[\phi]
=\int^{1}_{0}d\xi\int dt dx \Psi^{(\nu)\dagger}_{\sigma k\omega}(x,t,[\xi\phi])
\phi_{\sigma}(x,t)\Psi^{(\nu)}_{\sigma k\omega}(x,t,[\xi\phi])$,
$E^{(\pm)}_{k\omega}[0]=\omega\mp E_{k}$, and
$E_{k}=\sqrt{k^{2}+|\Delta|^{2}}$.

At $\phi_{R(L)\sigma}(x,t)=0$
(i.e., $V=0$), we have the solutions corresponding to the eigen-values
$E^{(\pm)}_{k\omega}[0]$, respectively,
\begin{eqnarray}
\Psi^{(\pm)}_{\sigma k\omega}(x,t)=\left(\frac{1}{TL}\right)^{1/2}
\left( \begin{array}{cc}
a^{(\pm)}_{k}e^{i\theta}\\ b^{(\pm)}_{k}e^{-i\theta} \end{array}\right)
e^{ikx-i\omega t}
\nonumber\end{eqnarray}
where $a^{(+)}_{k}=b^{(-)}_{k}=u_{k}$, $-b^{(+)}_{k}=a^{(-)}_{k}=
v_{k}$, $\Delta=\Delta^{(R)}+i\Delta^{(I)}$, and $\tan(2\theta)=\Delta^{(I)}
/\Delta^{(R)}$. The Hubbard-Stratonovich field
becomes a constant under this condition, $\Delta_{\sigma}(x,t)=\Delta$, and
the parameter $\Delta$ is determined by the self-consistent equation,
\begin{equation}
\Delta=-\frac{\lambda }{2L}\sum_{k}\frac{\Delta^{*}}{
\sqrt{k^{2}+|\Delta|^{2}}}
\label{27'}\end{equation}
which only has imaginary solution, $\Delta^{(R)}=0$, and
$\Delta^{(I)}\sim De^{-2\pi\hbar v_{F}/\lambda}$, 
where $D$ is the band-width of the system. The parameter $\Delta^{(I)}$ is a
charge gap parameter, and at $\lambda=0$ (i.e., $U=0$) it is zero. Therefore,
it can be taken as an order parameter of the Mott insulator.

Generally, we can choose the ansatz,
\begin{equation}
\Psi^{(\pm)}_{\sigma k\omega}(x,t,[\phi])=A_{k}\left(\frac{1}{TL}\right)^{1/2}
\left( \begin{array}{cc}
a^{(\pm)}_{k}e^{i\theta}e^{Q^{(\pm)}_{R\sigma}(x,t)}\\
b^{(\pm)}_{k}e^{-i\theta}
e^{Q^{(\pm)}_{L\sigma}(x,t)} \end{array}\right)
e^{ikx-i(\omega+\Sigma^{(\pm)}_{\sigma k}[\phi])t}
\label{28}\end{equation}
where $A_{k}\sim 1$ is a normalization constant, and the Hubbard-Stratonovich
field reads $\Delta_{\sigma}(x,t)=i\Delta^{(I)}e^{Q^{(-)}_{L-\sigma}(x,t)-
Q^{(-)}_{R-\sigma}(x,t)}$.
Substituting (\ref{28}) into the eigen-equation (\ref{27}), we can obtain,
\begin{eqnarray}
{\cal D}_{R}f^{(\pm)}_{R\sigma}+\displaystyle{ 
\frac{\Delta^{(I)}b^{(\pm)}_{k}}{a^{(\pm)}_{k}}
\left(e^{Q^{(-)}_{L-\sigma}-Q^{(-)}_{R-\sigma}}e^{
Q^{(\pm)}_{L\sigma}-Q^{(\pm)}_{R\sigma}} -1\right)} &=& 0 \label{210} \\
{\cal D}_{L}f^{(\pm)}_{L\sigma}+\displaystyle{ 
\frac{\Delta^{(I)}a^{(\pm)}_{k}}
{b^{(\pm)}_{k}}
\left(e^{Q^{(-)}_{R-\sigma}-Q^{(-)}_{L-\sigma}}e^{
Q^{(\pm)}_{R\sigma}-Q^{(\pm)}_{L\sigma}} -1\right)} &=& 0
\nonumber\end{eqnarray}
where $Q^{(\pm)}_{R(L)\sigma}(x,t)=Q^{0}_{R(L)\sigma}(x,t)
+f^{(\pm)}_{R(L)\sigma}(x,t)$.
These differential equations can be simplified to the Eikonal-type equations,
\begin{eqnarray}
&&\left\{ \begin{array}{ll}
\displaystyle{ ( {\cal D}_{L}{\cal D}_{R}-\frac{\Delta^{(I)}b^{(\pm)}_{k}}
{a^{(\pm)}_{k}}{\cal D}_{L}-\frac{\Delta^{(I)}a^{(\pm)}_{k}}{b^{(\pm)}_{k}}
{\cal D}_{R} )f^{(\pm)}_{R\sigma}} =& \displaystyle{
(\frac{\Delta^{(I)}b^{(\pm)}_{k}}{a^{(\pm)}_{k}}-
{\cal D}_{R}f^{(\pm)}_{R\sigma}
){\cal D}_{L}Q^{0}_{\sigma}+Z^{(\pm)}_{\sigma}}  \\
\displaystyle{ ( {\cal D}_{R}{\cal D}_{L}-\frac{\Delta^{(I)}b^{(\pm)}_{k}}
{a^{(\pm)}_{k}}{\cal D}_{L}-\frac{\Delta^{(I)}a^{(\pm)}_{k}}{b^{(\pm)}_{k}}
{\cal D}_{R} )f^{(\pm)}_{L\sigma}} =& \displaystyle{ -
(\frac{\Delta^{(I)}a^{(\pm)}_{k}}{b^{(\pm)}_{k}}-
{\cal D}_{L}f^{(\pm)}_{L\sigma})
{\cal D}_{R}Q^{0}_{\sigma}+\tilde{Z}^{(\pm)}_{\sigma}}\end{array}\right.
\label{210'} \\
&&\left\{  \begin{array}{ll}
\displaystyle{ Z^{(\pm)}_{\sigma}} &
 = \displaystyle{ \frac{\Delta^{(I)}b^{(\pm)}_{k}}{a^{(\pm)}_{k}} 
e^{-Q^{(\pm)}_{\sigma}}({\cal D}_{L}+{\cal D}_{L}f^{(\pm)}_{L\sigma}
)(e^{-Q^{(-)}_{-\sigma}}-1)} \\ & + \displaystyle{
(\Delta^{(I)})^{2}(e^{-Q^{(-)}_{-\sigma}}-1)
(e^{Q^{(-)}_{-\sigma}}-e^{-Q^{(\pm)}_{\sigma}})
-{\cal D}_{R}f^{(\pm)}_{R\sigma}
{\cal D}_{L}f^{(\pm)}_{R\sigma}} \\
\tilde{Z}^{(\pm)}_{\sigma} & 
= \displaystyle{
\frac{\Delta^{(I)}a^{(\pm)}_{k}}{b^{(\pm)}_{k}}
e^{Q^{(\pm)}_{\sigma}}({\cal D}_{R}-{\cal D}_{R}f^{(\pm)}_{R\sigma}
)(e^{Q^{(-)}_{-\sigma}}-1)} \\ & + \displaystyle{
(\Delta^{(I)})^{2}( 
e^{Q^{(-)}_{-\sigma}}-1)(e^{-Q^{(-)}_{-\sigma}}-
e^{Q^{(\pm)}_{\sigma}})
-{\cal D}_{R}f^{(\pm)}_{L\sigma}
{\cal D}_{L}f^{(\pm)}_{L\sigma}} 
\end{array}\right.
\nonumber\end{eqnarray}
where $Q^{0}_{\sigma}=Q^{0}_{R\sigma}-Q^{0}_{L\sigma}$, 
and $Q^{(\pm)}_{\sigma}=Q^{(\pm)}_{R\sigma}-Q^{(\pm)}_{L\sigma}$.

In order to solve these equations,we only keep
the linear terms of $f^{(\pm)}_{R(L)\sigma}(x,t)$ and 
$Q^{0}_{R(L)\sigma}$. Under this approximation, we obtain
the effective actions $S^{c}_{eff.}[\phi,\rho]$ and $S^{s}_{eff.}[\phi,\rho]$
of the spin and charge parts, respectively,
\begin{eqnarray}
S^{c(s)}_{eff.}[\phi,\rho] &=& \displaystyle{ \frac{1}{TL}\sum_{q,\Omega}
\left\{ -\frac{q}{8\pi}\frac{\Omega+q}{D_{c(s)}}|\phi_{Rc(s)}(q,\Omega)|^{2}
+\frac{q}{8\pi}\frac{\Omega-q}{D_{c(s)}}|\phi_{Lc(s)}(q,\Omega)|^{2}\right.}
\label{211} \\
&-& \displaystyle{ \frac{1}{2}\left[\phi_{Rc(s)}(-q,-\Omega)
\rho_{Rc(s)}(q,\Omega)+\phi_{Lc(s)}(-q,-\Omega)\rho_{Lc(s)}(q,\Omega)\right]} 
\nonumber \\
&\mp& \displaystyle{ \left.
\frac{V}{4}|\rho_{Rc(s)}(q,\Omega)+\rho_{Lc(s)}(q,\Omega)|^{2}\right\}}
\nonumber\end{eqnarray}
where $D_{c(s)}=\Omega^{2}-q^{2}-(\Delta^{(I)}\pm\Delta^{(I)})^{2}$,
and $\chi_{c(s)}=\chi_{\uparrow}\pm \chi_{\downarrow}$ where 
$\chi=\phi_{R(L)},\rho_{R(L)}$. 
Integrating out the auxiliary fields $\phi_{Rc(s)}(q,\Omega)$ and
$\phi_{Lc(s)}(q,\Omega)$, we obtain the effective actions,
\begin{eqnarray}
S^{c(s)}_{eff.}[\rho] &=& \displaystyle{ \frac{1}{TL}\sum_{q,\Omega}
\rho^{T}_{c(s)}(-q,-\Omega)X_{c(s)}(q,\Omega)\rho_{c(s)}(q,\Omega)} 
\nonumber \\
X_{c(s)}(q,\Omega) &=& \displaystyle{ A_{c(s)}\left( \begin{array}{cc}
q(\Omega-q)\mp\displaystyle{\frac{V}{4A_{c(s)}}, \;\;\;\; \mp\frac{V}
{4A_{c(s)}}}\\ \mp\displaystyle{ \frac{V}{4A_{c(s)}}, \;\;\;\; 
-q(\Omega+q)\mp\frac{V}{4A_{c(s)}}} \end{array}\right)}
\label{212}\end{eqnarray}
where $\rho^{T}_{c(s)}=(\rho_{Rc(s)},\rho_{Lc(s)})$, $A_{c(s)}=-\frac{
\pi D_{c(s)}}{2}(q^{4}-q^{2}\Omega^{2})^{-1}$. 
With these effective action, we can obtain the gaps in the spin and charge
collective excitation spectrum, respectively,
\begin{equation}
\left\{ \begin{array}{ll}
\Delta_{s}= & 0 \\
\Delta_{c}= & 2\Delta^{(I)} \end{array}\right.
\label{213}\end{equation}
At zero-order approximation (i.e., $V=0$), the parameter $\Delta^{(I)}$ is
determined by equation (\ref{27'}), and it increases with the electron
interaction strength $\lambda$ ($=aU$). However, the parameter
$\Delta^{(I)}$ can be more rigorously determined by taking the minimum of the
ground state energy of the system, where the ground state energy reads,
\begin{equation}
E_{g}=\frac{2(\Delta^{(I)})^{2}}{\lambda}-i Tr\ln\left(\hat{M}_{0}\right)
+\frac{i}{2} Tr\ln\left(\frac{D_{c}(V)}{4V D_{c}}\right)
\label{214}\end{equation}
where $D_{c}(V)=\Omega^{2}-(1+V/\pi)q^{2}-4(\Delta^{(I)})^{2}$. By simple
calculation, we can obtain,
\begin{eqnarray}
\Delta^{(I)} &=& \displaystyle{ \frac{\lambda\Delta^{(I)}}{\pi}
\ln\left(\frac{D+\sqrt{D^{2}+(\Delta^{(I)})^{2}}}{\Delta^{(I)}}\right)}
\nonumber \\
&-& \displaystyle{ \frac{\lambda\Delta^{(I)}}{2\pi}\frac{1}{\sqrt{1+V/\pi}}
\ln\left(\frac{\sqrt{1+V/\pi}D+\sqrt{(1+V/\pi)D^{2}+(\Delta^{(I)})^{2}}}
{\Delta^{(I)}}\right)}
\label{215}\end{eqnarray}
At $V=0$, it reduces into the equation (\ref{27'}). We now can approximately
take $\lambda=V$ in the low energy region, and use equations (\ref{215}) and
(\ref{213}) to
determine the order parameter $\Delta^{(I)}$ of the Mott insulator
and the charge collective excitation gap, respectively, where the only
parameter is the band-width $D$ of electrons.
This result is qualititively consistent with the exact solution of the Hubbard
model at half-filling, where the charge gap increases with the on-site
Coulomb interaction strength of electrons\cite{12}.

\section{phase transition from band insulator to Mott-type insulator}

In this section, we study the influence of strong electron correlation
on the spin and charge collective excitation gaps with a prototype
one-dimensional model\cite{9'}. This model can be used to qualititively explain
the origin of ferroelecticity of some transition metal oxides, because of a
spontaneous lattice dimerization in the strong electron correlation regime. 
By analytical calculations of the spin and charge gaps, we can clearly 
demonstrate that the system has a quantum critical point defined by
$\Delta_{c}=0$ at the interaction strength $V=V_{T}(\Delta_{0})$, where the
order parameter $\Delta^{(I)}$ of Mott insulator and dimerization parameter
$u$ turns on, and increase with $V$ as $V>V_{T}(\Delta_{0})$, and the order
parameter $\Delta^{(R)}$ of band insulator goes to a constant as 
$V>V_{T}(\Delta_{0})$. This quantum critical point represents the phase 
transition from band insulator to Mott-type insulator, and as $\Delta_{0}=0$,
it reduces into the usual metal-Mott insulator transition of the Hubbard
model at half-filling. Another quantum critical point defined by $\Delta_{s}=0$
is reached as $\Delta^{(I)}\rightarrow\infty$ asymptotically for a finite
$\Delta_{0}$. Qualititively, the phenomenal results of Ref.\cite{13} are 
consistent with our calculations.

The Hamiltonian of this prototype one-dimensional model reads,
\begin{equation}
H=\sum_{i\sigma}\left\{-t\left(c^{\dagger}_{i\sigma}c_{i+1\sigma}+
c^{\dagger}_{i+1\sigma}c_{i\sigma}\right)+\Delta_{0}(-1)^{i}n_{i\sigma}
\right\}+U\sum_{i}n_{i\uparrow}n_{i\downarrow}
\label{31}\end{equation}
where $c_{i\sigma}(c^{\dagger}_{i\sigma})$ is the electron annihilation
(creation) operator with spin $\sigma$ ($=\uparrow,\downarrow$)
at site $i$, and $n_{i\sigma}=c^{\dagger}_{i\sigma}c_{i\sigma}$ is the electron
density operator. The odd and even sites represent oxygen atoms (O) 
and a generic cation (C), respectively, with the energy difference,
$E_{C}-E_{O}=2\Delta_{0}$. It can be easily seen that the Hamiltonian 
(\ref{31})
describes different low energy physical behavior for the on-site
Coulomb interaction $U\rightarrow 0$ and $U\rightarrow\infty$, respectively.
At $U=0$, it represents a simple band insulator where there is an energy gap
$2\Delta_{0}$ in the electron excitation spectrum. 
However, as $U\gg\Delta_{0}$, it
represents a Mott insulator where charge density excitation spectrum has a
gap and spin density excitation spectrum is gapless.
Therefore, there is a transition (mixed-valence) region 
between the band insulator and Mott-type
insulator as $U\sim 2\Delta_{0}$.
The finite-size simulations\cite{9',14,15} suggest that
the mixed-valence region is accompanied by a spontaneous dimerized phase
which turns on electron-phonon interaction,
\begin{equation}
H_{ep}=-u\sum_{i\sigma}(-1)^{i}\left(c^{\dagger}_{i\sigma}c_{i+1\sigma}
+c^{\dagger}_{i+1\sigma}c_{i\sigma}\right)
\label{32}\end{equation}
where $u$ is the dimerization parameter.

In low energy region, and with the linearization of electron spectrum
near the Fermi levels $\pm k_{F}$,
the Hamiltonian (\ref{31}) can be written as in the 
continuum limit,
\begin{eqnarray}
H &=& \displaystyle{ -i\hbar v_{F}\sum_{\sigma}\int dx \left[
\psi^{\dagger}_{R\sigma}(x)\partial_{x}\psi_{R\sigma}(x)-
\psi^{\dagger}_{L\sigma}(x)\partial_{x}\psi_{L\sigma}(x)\right]} \nonumber \\
&+& \displaystyle{ V\int dx\left[\rho_{R\uparrow}(x)+\rho_{L\uparrow}(x)
\right]\left[\rho_{R\downarrow}(x)+\rho_{L\downarrow}(x)\right]} \label{33} \\
&-& \displaystyle{ \Delta_{0}\sum_{\sigma}\int dx\left[
\psi^{\dagger}_{R\sigma}(x)\psi_{L\sigma}(x)+\psi^{\dagger}_{L\sigma}(x)
\psi_{R\sigma}(x)\right]} \nonumber \\
&+& \displaystyle{\lambda\int dx\left[\psi^{\dagger}_{R\uparrow}(x)
\psi_{L\uparrow}(x)\psi^{\dagger}_{R\downarrow}(x)\psi_{L\downarrow}(x)+
\psi^{\dagger}_{L\uparrow}(x)
\psi_{R\uparrow}(x)\psi^{\dagger}_{L\downarrow}(x)\psi_{R\downarrow}(x)\right]}
\nonumber\end{eqnarray}
where $\psi_{R\sigma}(x)$ ($\psi_{L\sigma}(x)$) is the right (left) moving
electron field with spin $\sigma$, 
$\rho_{R(L)\sigma}(x)=\psi^{\dagger}_{R(L)\sigma}(x)
\psi_{R(L)\sigma}(x)$ is the electron density field, $V=aU$, and $\lambda
=aU$, where $a$ is the lattice constant. Here we have used $\lambda$ to
present the coefficient of the Umklapp scattering which is relevant for 
repulsive electron Coulomb interaction ($U>0$), therefore, in the low
energy region renormalized $\lambda$ is different from $aU$, and neglected the
backward scattering term which is irrelevant for the repulsive interaction
$V>0$. The spontaneous electron-phonon coupling term (\ref{32}) reads in
the continuum limit,
\begin{equation}
H_{ep}=iu\sum_{\sigma}\int dx \left[\psi^{\dagger}_{R\sigma}(x)
\psi_{L\sigma}(x)-\psi^{\dagger}_{L\sigma}(x)\psi_{R\sigma}(x)\right]
\label{34}\end{equation}
The dimerization parameter $u$ can be determined by taking the minimum 
of ground state energy after including static tension energy of the lattices.

With auxiliary fields $\phi_{R\sigma}(x,t)$ and $\phi_{L\sigma}(x,t)$
which introduce the constraints 
$\rho_{R\sigma}(x,t)=\psi^{\dagger}_{R\sigma}(x,t)\psi_{R\sigma}(x,t)$ and 
$\rho_{L\sigma}(x,t)=\psi^{\dagger}_{L\sigma}(x,t)
\psi_{L\sigma}(x,t)$, respectively, and the Hubbard-Stratonovich field
$\Delta_{-\sigma}(x,t)=\psi^{\dagger}_{R\sigma}(x,t)\psi_{L\sigma}(x,t)$,
the action of the system without the lattice dimerization can be written as,
\begin{eqnarray}
S &=& \displaystyle{ \sum_{\sigma}\int dt dx\left\{\Psi^{\dagger}_{\sigma}(x,t)
\hat{M}_{\sigma}(x,t)\Psi_{\sigma}(x,t)\right.} \nonumber \\
&-& \displaystyle{ \left.\phi_{R\sigma}(x,t)\rho_{R\sigma}(x,t)
-\phi_{L\sigma}(x,t)\rho_{L\sigma}(x,t)\right\}} \label{35} \\
&+& \displaystyle{ \int dt dx\left\{\frac{1}{\lambda}\left[
\Delta_{\uparrow}(x,t)\Delta_{\downarrow}(x,t)+\Delta^{*}_{\uparrow}(x,t)
\Delta^{*}_{\downarrow}(x,t)\right]\right.} \nonumber \\
&-& \displaystyle{ \left.V\left[\rho_{R\uparrow}(x,t)+
\rho_{L\uparrow}(x,t)\right]\left[\rho_{R\downarrow}(x,t)+
\rho_{L\downarrow}(x,t)\right]\right\}}
\nonumber\end{eqnarray}
where $\hat{M}_{\sigma}(x,t)=\hat{M}_{0}+\phi_{\sigma}(x,t)$,
$\Psi^{\dagger}_{\sigma}(x,t)=\left(\psi^{\dagger}_{R\sigma}(x,t),
\psi^{\dagger}_{L\sigma}(x,t)\right)$, and 
\begin{eqnarray}
\hat{M}_{0}=\left( \begin{array}{cc}
{\cal D}_{R}, \;\; \Delta_{0}+\Delta\\
\Delta_{0}+\Delta^{*}, \;\; {\cal D}_{L} \end{array}\right), \;\;\;\;
\phi_{\sigma}(x,t)=\left( \begin{array}{cc}
\phi_{R\sigma}(x,t), \;\; \Delta_{\sigma}(x,t)-\Delta\\
\Delta^{*}_{\sigma}(x,t)-\Delta^{*}, \;\; \phi_{L\sigma}(x,t)\end{array}
\right)\nonumber\end{eqnarray}
where ${\cal D}_{R(L)}=i\partial_{t}\pm i\partial_{x}$ (choosing $\hbar=
v_{F}=1$), and $\Delta$ is a parameter (see below). Integrating out the
electron field $\Psi_{\sigma}(x,t)$, we obtain an effective potential,
$W[\phi]=i Tr\ln(\hat{M}_{\sigma})$ which can be represented,
\begin{eqnarray}
W[\phi] &=& \displaystyle{ i\sum_{\sigma}
\int^{1}_{0} d\xi\int dt dx \; tr\left(\phi_{\sigma}G_{\sigma}(x,t;x',t',
[\xi\phi])\right)|_{\stackrel{t'\rightarrow t}{x'\rightarrow x}}} \nonumber \\
G_{\sigma}(x,t;x',t',[\phi]) &=& \displaystyle{ \sum_{\nu}
\sum_{k,\omega}\frac{1}{E^{(\nu)}_{\sigma k\omega}[\phi]}
\Psi^{(\nu)}_{\sigma k\omega}(x,t,[\phi])\Psi^{(\nu)\dagger}_{\sigma k\omega}
(x',t',[\phi])}
\label{36}\end{eqnarray}
where the eigen-functional $\Psi^{(\nu)}_{\sigma k\omega}(x,t,[\phi])$ 
satisfies the eigen-equation,
\begin{equation}
\hat{M}_{\sigma}(x,t)\Psi^{(\nu)}_{\sigma k\omega}(x,t,[\phi])
=E^{(\nu)}_{\sigma k\omega}[\phi]\Psi^{(\nu)}_{\sigma k\omega}(x,t,[\phi])
\label{37}\end{equation}
Using Hellmann-Feynman theorem, the eigen-value $E^{(\nu)}_{\sigma k\omega}
[\phi]$ reads $E^{(\nu)}_{\sigma k\omega}[\phi]=E^{(\nu)}_{k\omega}[0]+
\Sigma^{(\nu)}_{\sigma k}[\phi]$, where $\Sigma^{(\nu)}_{\sigma k}[\phi]
=\int^{1}_{0}d\xi\int dt dx \Psi^{(\nu)\dagger}_{\sigma k\omega}(x,t,[\xi\phi])
\phi_{\sigma}(x,t)\Psi^{(\nu)}_{\sigma k\omega}(x,t,[\xi\phi])$,
independent of $\omega$.

In order to solve the eigen-equation (\ref{37}), we can choose the ansatz,
\begin{equation}
\Psi^{(\pm)}_{\sigma k\omega}(x,t,[\phi])=A_{k}\left(\frac{1}{TL}\right)^{1/2}
\left( \begin{array}{cc}
\alpha^{(\pm)}_{k}e^{i\tilde{\theta}}e^{Q^{(\pm)}_{R\sigma}(x,t)}\\
\beta^{(\pm)}_{k}e^{-i\tilde{\theta}}
e^{Q^{(\pm)}_{L\sigma}(x,t)} \end{array}\right)
e^{ikx-i(\omega+\Sigma^{(\pm)}_{\sigma k}[\phi])t}
\label{38}\end{equation}
where $\tan(2\tilde{\theta})=\Delta^{(I)}/(\Delta_{0}+\Delta^{(R)})$,
the eigen-value is $E^{(\pm)}_{\sigma k\omega}[\phi]=\omega\mp E_{k}+
\Sigma^{(\pm)}_{\sigma k}[\phi]$, $E_{k}=\sqrt{k^{2}+|\Delta_{0}+
\Delta|^{2}}$, $\alpha^{(+)}_{k}=\beta^{(-)}_{k}=
\sqrt{\frac{1}{2}(1+k/E_{k})}$, $-\beta^{(+)}_{k}=\alpha^{(-)}_{k}=
\sqrt{\frac{1}{2}(1-k/E_{k})}$, $A_{k}\sim 1$ is a normalization constant,
and $\Delta_{\sigma}(x,t)=\Delta
e^{Q^{(-)}_{L-\sigma}(x,t)-Q^{(-)}_{R-\sigma}(x,t)}$, 
where $\Delta=\Delta^{(R)}+i\Delta^{(I)}$ is
determined by the self-consistent equation,
\begin{eqnarray}
\Delta^{(R)} &=& \displaystyle{-\frac{\lambda}{2L}
\sum_{k}\frac{\Delta_{0}+\Delta^{(R)}}{E_{k}}} \label{39} \\
\Delta^{(I)} &=& \displaystyle{ \frac{\lambda}{2L}
\sum_{k}\frac{\Delta^{(I)}}{E_{k}}}
\nonumber\end{eqnarray}
For a finite $\Delta_{0}$, the real part $\Delta^{(R)}$ has a non-zero 
solution ($\Delta^{(R)}<0$), and $\Delta^{(I)}=0$ for $V<V_{T}(\Delta_{0})$,
and at $\Delta_{0}=0$ 
it only has a zero solution $\Delta^{(R)}=0$, where the system reduces to
the usual Hubbard model at half-filling. Therefore, the parameter 
$\Delta^{(R)}$ characterizes the band insulator, and $\Delta^{(I)}$
characterizes the Mott insulator. The parameters $\Delta^{(R)}$ and
$\Delta^{(I)}$ can be taken as the order parameters of band insulator and
Mott insulator, respectively.

Substituting (\ref{38}) into the eigen-equation (\ref{37}), we can obtain,
\begin{eqnarray}
{\cal D}_{R}f^{(\pm)}_{R\sigma}+\displaystyle{ \frac{m\beta^{(\pm)}_{k}}{
\alpha^{(\pm)}_{k}}\left(e^{Q^{(\pm)}_{L\sigma}-Q^{(\pm)}_{R\sigma}}-1
\right)+\frac{\Delta\beta^{(\pm)}_{k}e^{-i2\tilde{\theta}}}{\alpha^{(\pm)}_{k}}
\left(e^{Q^{(-)}_{L-\sigma}-Q^{(-)}_{R-\sigma}}-1\right)e^{
Q^{(\pm)}_{L\sigma}-Q^{(\pm)}_{R\sigma}}} &=& 0 \label{310} \\
{\cal D}_{L}f^{(\pm)}_{L\sigma}+\displaystyle{ \frac{m\alpha^{(\pm)}_{k}}{
\beta^{(\pm)}_{k}} \left(e^{Q^{(\pm)}_{R\sigma}-Q^{(\pm)}_{L\sigma}}-1
\right)+\frac{\Delta^{*}\alpha^{(\pm)}_{k}e^{i2\tilde{\theta}}}
{\beta^{(\pm)}_{k}}
\left(e^{Q^{(-)}_{R-\sigma}-Q^{(-)}_{L-\sigma}}-1\right)e^{
Q^{(\pm)}_{R\sigma}-Q^{(\pm)}_{L\sigma}}} &=& 0
\nonumber\end{eqnarray}
where $m=[(\Delta_{0}+\Delta^{(R)})^{2}+(\Delta^{(I)})^{2}]^{1/2}$, and
$Q^{(\pm)}_{R(L)\sigma}(x,t)=Q^{0}_{R(L)\sigma}(x,t)
+f^{(\pm)}_{R(L)\sigma}(x,t)$.
These differential equations can be simplified to the Eikonal-type equations,
\begin{eqnarray}
&&\left\{ \begin{array}{ll}
\displaystyle{ \left( {\cal D}_{L}{\cal D}_{R}-\frac{m\beta^{(\pm)}_{k}}
{\alpha^{(\pm)}_{k}}{\cal D}_{L}-\frac{m\alpha^{(\pm)}_{k}}{\beta^{(\pm)}_{k}}
{\cal D}_{R} \right)f^{(\pm)}_{R\sigma}} =& \displaystyle{
\left(\frac{m\beta^{(\pm)}_{k}}{\alpha^{(\pm)}_{k}}-
{\cal D}_{R}f^{(\pm)}_{R\sigma}
\right){\cal D}_{L}Q^{0}_{\sigma}+Z^{(\pm)}_{\sigma}}  \\
\displaystyle{ \left( {\cal D}_{R}{\cal D}_{L}-\frac{m\beta^{(\pm)}_{k}}
{\alpha^{(\pm)}_{k}}{\cal D}_{L}-\frac{m\alpha^{(\pm)}_{k}}{\beta^{(\pm)}_{k}}
{\cal D}_{R} \right)f^{(\pm)}_{L\sigma}} =& \displaystyle{ -
\left(\frac{m\alpha^{(\pm)}_{k}}{\beta^{(\pm)}_{k}}-
{\cal D}_{L}f^{(\pm)}_{L\sigma}
\right){\cal D}_{R}Q^{0}_{\sigma}+\tilde{Z}^{(\pm)}_{\sigma}}\end{array}\right.
\label{310'} \\
&&\left\{  \begin{array}{ll}
\displaystyle{ Z^{(\pm)}_{\sigma}} &
 = \displaystyle{ 
e^{-f^{(\pm)}_{R\sigma}}\left({\cal D}_{L}+{\cal D}_{L}Q^{0}_{\sigma}-
\frac{m\alpha^{(\pm)}_{k}}{\beta^{(\pm)}_{k}}\right)
F^{(\pm)}_{\sigma}-\frac{m\beta^{(\pm)}_{k}}{\alpha^{(\pm)}_{k}}
e^{-Q^{0}_{\sigma}-f^{(\pm)}_{R\sigma}}
\tilde{F}^{(\pm)}_{\sigma}
-{\cal D}_{R}f^{(\pm)}_{R\sigma}
{\cal D}_{L}f^{(\pm)}_{R\sigma}} \\
\tilde{Z}^{(\pm)}_{\sigma} & = \displaystyle{
e^{-f^{(\pm)}_{L\sigma}}\left({\cal D}_{R}-{\cal D}_{R}Q^{0}_{\sigma}-
\frac{m\beta^{(\pm)}_{k}}{\alpha^{(\pm)}_{k}}\right)
\tilde{F}^{(\pm)}_{\sigma}-\frac{m\alpha^{(\pm)}_{k}}{\beta^{(\pm)}_{k}} 
e^{Q^{0}_{\sigma}-f^{(\pm)}_{L\sigma}}F^{(\pm)}_{\sigma}
-{\cal D}_{R}f^{(\pm)}_{L\sigma}
{\cal D}_{L}f^{(\pm)}_{L\sigma}} \end{array}\right.
\nonumber\end{eqnarray}
where $Q^{0}_{\sigma}=Q^{0}_{R\sigma}-Q^{0}_{L\sigma}$, $F^{(\pm)}_{\sigma}
=\Delta\beta^{(\pm)}_{k}e^{-i2\tilde{\theta}}e^{-Q^{0}_{\sigma}+f^{(\pm)}
_{L\sigma}}(e^{Q^{(-)}_{L-\sigma}-Q^{(-)}_{R-\sigma}}-1)/\alpha^{(\pm)}_{k}$,
and $\tilde{F}^{(\pm)}_{\sigma}=\Delta^{*}\alpha^{(\pm)}_{k}e^{i2
\tilde{\theta}}e^{Q^{0}_{\sigma}+f^{(\pm)}_{R\sigma}}(e^{Q^{(-)}_{R-\sigma}
-Q^{(-)}_{L-\sigma}}-1)/\beta^{(\pm)}_{k}$.
At $\Delta=0$, it is reduced to the same Eikonal-type equations as
that in the equation (\ref{7}). 

In order to solve these Eikonal-type equations,we only keep
the linear terms of $f^{(\pm)}_{R(L)\sigma}(x,t)$ and 
$Q^{0}_{R(L)\sigma}$. Under this approximation, we obtain
the effective actions $S^{c}_{eff.}[\phi,\rho]$ and $S^{s}_{eff.}[\phi,\rho]$
of the spin and charge parts, respectively,
\begin{eqnarray}
S^{c(s)}_{eff.}[\phi,\rho] &=& \displaystyle{ \frac{1}{TL}\sum_{q,\Omega}
\left\{ -\frac{q}{8\pi}\frac{\Omega+q}{D_{c(s)}}|\phi_{Rc(s)}(q,\Omega)|^{2}
+\frac{q}{8\pi}\frac{\Omega-q}{D_{c(s)}}|\phi_{Lc(s)}(q,\Omega)|^{2}\right.}
\label{311} \\
&-& \displaystyle{ \frac{1}{2}\left[\phi_{Rc(s)}(-q,-\Omega)
\rho_{Rc(s)}(q,\Omega)+\phi_{Lc(s)}(-q,-\Omega)\rho_{Lc(s)}(q,\Omega)\right]} 
\nonumber \\
&\mp& \displaystyle{ \left.
\frac{V}{4}|\rho_{Rc(s)}(q,\Omega)+\rho_{Lc(s)}(q,\Omega)|^{2}\right\}}
\nonumber\end{eqnarray}
where $D_{c(s)}=\Omega^{2}-q^{2}-(m\pm\tilde{\Delta})^{2}$,
$\tilde{\Delta}=\sqrt{(\Delta^{(R)})^{2}+(\Delta^{(I)})^{2}}\cos\left(2(
\tilde{\theta}-\theta)\right)$, $\tan(2\theta)=\Delta^{(I)}/\Delta^{(R)}$, and
$\chi_{c(s)}=\chi_{\uparrow}\pm \chi_{\downarrow}$ where 
$\chi=\phi_{R(L)},\rho_{R(L)}$. 
Integrating out the auxiliary fields $\phi_{Rc(s)}(q,\Omega)$ and
$\phi_{Lc(s)}(q,\Omega)$, we obtain the effective actions,
\begin{eqnarray}
S^{c(s)}_{eff.}[\rho] &=& \displaystyle{ \frac{1}{TL}\sum_{q,\Omega}
\rho^{T}_{c(s)}(-q,-\Omega)X_{c(s)}(q,\Omega)\rho_{c(s)}(q,\Omega)} 
\nonumber \\
X_{c(s)}(q,\Omega) &=& \displaystyle{ A_{c(s)}\left( \begin{array}{cc}
q(\Omega-q)\mp\displaystyle{\frac{V}{4A_{c(s)}}, \;\;\;\; \mp\frac{V}
{4A_{c(s)}}}\\ \mp\displaystyle{ \frac{V}{4A_{c(s)}}, \;\;\;\; 
-q(\Omega+q)\mp\frac{V}{4A_{c(s)}}} \end{array}\right)}
\label{312}\end{eqnarray}
where $\rho^{T}_{c(s)}=(\rho_{Rc(s)},\rho_{Lc(s)})$, $A_{c(s)}=-\frac{
\pi D_{c(s)}}{2}(q^{4}-q^{2}\Omega^{2})^{-1}$. 

With the effective actions $S^{s}_{eff.}[\rho]$ and $S^{c}_{eff.}
[\rho]$ (\ref{312}), we can obtain the gaps in the spin and charge collective
excitation spectrums, respectively,
\begin{equation}
\left\{ \begin{array}{ll}
\Delta^{2}_{s}= & \displaystyle{(m-\tilde{\Delta})^{2}} \\
\Delta^{2}_{c}= & \displaystyle{(m+\tilde{\Delta})^{2}} \end{array}\right.
\label{313}\end{equation} 
At $\Delta_{0}=0$, it reduces to
the well-known result of the Hubbard model at half-filling that $\Delta_{s}=0$
and $\Delta^{2}_{c}=4(\Delta^{(I)})^{2}$, where
the charge gap is increased with the
electron interaction strength $U$. This can be seen from the 
self-consistent equation of $\Delta^{(I)}$ (\ref{39}) where $\Delta^{(I)}$
increases with $\lambda$. Generally, the parameter $\lambda$ can be determined
by the minimum of the ground state energy, and it is a function of $V$,
$\lambda(V)$. However, we can approximately take $\lambda=V$ which does not
qualitatively change the behavior of the spin and charge gaps determined by
equations (\ref{39}) and (\ref{313}), and further simplifies our calculation.
Equation (\ref{313}) clearly shows that the system has two quantum critical
points defined by the charge and spin gaps, one is at $\Delta_{c}=0$, and 
another is at $\Delta_{s}=0$. The physical behavior around the quantum
critical points can be understood by solving the self-consistent equation
(\ref{39}). By simple calculation, we can obtain the following solutions,
\begin{eqnarray}
\Delta^{(I)} &=& \displaystyle{\left\{ \begin{array}{ll}
0, & \;\; V\leq V_{T}(\Delta_{0})\\
\displaystyle{\left[\left(\frac{2De^{2\pi\hbar v_{F}/V}}
{e^{4\pi\hbar v_{F}/V}-1}\right)^{2}-\frac{1}{4}\Delta^{2}_{0}\right]^{1/2}}, &
\;\; V>V_{T}(\Delta_{0}) \end{array}\right.} \nonumber \\ 
\Delta^{(R)} &=& \displaystyle{\left\{ \begin{array}{ll}
\displaystyle{\Delta^{(R)}(V)}, & \;\; V\leq V_{T}(\Delta_{0})\\
\displaystyle{-\frac{1}{2}\Delta_{0}}, & \;\; V>V_{T}(\Delta_{0}) 
\end{array}\right.}
\label{314}\end{eqnarray}
where $V_{T}(\Delta_{0})=2\pi\hbar v_{F}/\ln[2(D+\sqrt{D^{2}+
\Delta^{2}_{0}/4})/\Delta_{0}]$, and $\Delta^{(R)}(V)$ is determined by 
equation (\ref{39}) 
with $\Delta^{(I)}=0$. The interaction $V=V_{T}(\Delta_{0})$
corresponds to the quantum critical point $\Delta_{c}=0$. It means that
there exists a phase transition at $\Delta_{c}=0$ where the
parameter $\Delta_{c}$ shows different behavior on the two sides of the point 
$V=V_{T}(\Delta_{0})$. As $V<V_{T}(\Delta_{0})$, the parameter
$\Delta^{(I)}$ is zero, and $\Delta^{(R)}$ decreases with increasing
$V$. In this case, the parameter $\Delta$ is real.
On the other hand, as $V>V_{T}(\Delta_{0})$, 
the parameter $\Delta^{(R)}$ becomes
a constant, and $\Delta^{(I)}$ is not zero and increases with $V$. 
The parameter $\Delta$
becomes a complex quantity. For the strong 
interaction $V>V_{T}(\Delta_{0})$, the charge and spin gaps can be written as,
respectively,
\begin{equation}
\left\{ \begin{array}{ll}
\Delta_{c}= & \displaystyle{ \frac{4(\Delta^{(I)})^{2}}{\sqrt{\Delta^{2}_{0}+
4(\Delta^{(I)})^{2}}}}\\
\Delta_{s}= & \displaystyle{ \frac{\Delta^{2}_{0}}{\sqrt{\Delta^{2}_{0}+
4(\Delta^{(I)})^{2}}}  } \end{array}\right.
\label{315}\end{equation}
It is noted that for weack interaction $V<V_{T}(\Delta_{0})$ 
the spin gap retain invariant,
and the quantum critical point $\Delta_{s}=0$ corresponds to $\Delta^{(I)}
\rightarrow\infty$ (i.e., $V,\; D\rightarrow\infty$). In fact, this is due to
our simple approximations that the self-consistent equation (\ref{39})
is obtained at $\phi_{\sigma}(x,t)=0$ (i.e., $V=0$), it does not include the
high-order correction of $V$. 
However, our results are qualitatively correct
after including this influence.

Including the electron-phonon interaction, the equation
(\ref{313}) keeps invariant, but the parameters $m$, $\tilde{\theta}$ and
the self-consistent equation of $\Delta$ are changed as,
\begin{eqnarray}
m &=& \displaystyle{ [(\Delta_{0}+\Delta^{(R)})^{2}+(\Delta^{(I)}-
u)^{2}]^{1/2}, \;\;\;\; \tan(2\tilde{\theta})=\frac{\Delta^{(I)}-u}{
\Delta_{0}+\Delta^{(R)}}} \nonumber \\
\Delta^{(R)} &=& \displaystyle{ -\frac{\lambda}{2L}\sum_{k}
\frac{\Delta_{0}+\Delta^{(R)}}{E_{k}}}, \;\;\;\;
\Delta^{(I)} = \displaystyle{ \frac{\lambda}{2L}\sum_{k}
\frac{\Delta^{(I)}-u}{E_{k}}}
\label{316}\end{eqnarray}
where $E_{k}=[k^{2}+m^{2}]^{1/2}$, $\tilde{\Delta}=|\Delta|\cos\left(2(\theta-
\tilde{\theta})\right)$,
and $\tan(2\theta)=\Delta^{(I)}/\Delta^{(R)}$.
Here we only consider the static uniform dimerization of the lattices.
Including the static tension energy of the lattices $\frac{1}{2}\alpha
u^{2}$, the dimerization parameter $u$ can be determined by the self-consistent
equation,
\begin{eqnarray}
-\alpha u &=& \displaystyle{ \frac{2}{L}\sum_{k}\frac{\Delta^{(I)}-u}{E_{k}}
-\frac{m+\tilde{\Delta}}{2L}\left(\frac{\Delta^{(I)}-u}{m}-z\right)
\sum_{q}\left( \frac{1}{E^{c}_{q}(V)}-\frac{1}{E^{c}_{q}}
\right)} \nonumber \\
&-& \displaystyle{ \frac{m-\tilde{\Delta}}{2L}\left( \frac{\Delta^{(I)}
-u}{m}+z\right)\sum_{q}\left(\frac{1}{E^{s}_{q}(V)}-\frac{1}{E^{s}_{q}}
\right)}
\label{317}\end{eqnarray}
where $z=|\Delta|
\sin\left(2(\tilde{\theta}-\theta)\right)\cos^{2}(2\tilde{\theta})/(\Delta_{0}
+\Delta^{(R)})$, $E^{c(s)}_{q}(V)=[(1\pm V/\pi)q^{2}+\Delta^{2}_{c(s)}]^{1/2}$,
and $E^{c(s)}_{q}=[q^{2}+\Delta^{2}_{c(s)}]^{1/2}$.
This equation is derived from the minimum of the ground state energy,
$\partial_{u}E_{g}[u]=0$. According to equations (\ref{316}) and (\ref{317}), 
we see that the dimerization parameter $u$, similar to $\Delta^{(I)}$, is zero 
for weack interaction $V<V_{T}(\Delta_{0})$, and turns on at 
$V=V_{T}(\Delta_{0})$. However, for strong interaction
$V>V_{T}(\Delta_{0})$, the parameter $\Delta^{(R)}$ basically keeps a 
constant, because the dimerization parameter $u$ is a small quantity,
and it goes to zero as $V\gg V_{T}(\Delta_{0})$.

In usual transition metal oxides, there is the strong hybridization between
d-orbit electrons of cation and p-orbit electrons of oxygen atom. Therefore,
at enough strong electron correlation, there spontaneously takes place the
lattice dimerization and deformation-induced charge transfer, which is
possibly the origin of ferroelectricity of the transition metal oxides.
Our calculations demonstrate that in the strong interaction region
$V>V_{T}(\Delta_{0})$, the system has the spontaneous lattice dimerization,
which is consistent with the previous numerical calculations\cite{9',14,15}.

\section{discussion and conclusion}

In general, for a one-dimensional strongly correlated electron system, if
there is only low energy excitation modes with small momentum near the
Fermi levels $\pm k_{F}$, it shows the Luttinger liquid behavior. If there are
disorder impurities which induces the new low energy excitation modes with
large momentum transfer between two Fermi levels $\pm k_{F}$, the electrons
are localized, and the system becomes an insulator. At electron half-filling,
there appears the Umklapp scattering which represents the low energy excitation
modes with large momentum $4k_{F}$ transfer between the Fermi levels
$\pm k_{F}$, the system becomes a Mott insulator. Therefore, the low energy
excitation with large momentum transfer between the Fermi levels $\pm k_{F}$
(i.e., crossover excitation)
makes the system become the insulator for repulsive electron interaction.

Usual perturbation methods fail to treat such the system, because the crossover
excitation term(s) is relevant, and determines the low energy behavior of
the system. It is very desirable to find a new method which can exactly and
effectively treat this kind of systems.
The eigen-functional bosonization is a good candidate, because it has more
advantages than usual perturbation theory: a). Without the crossover
excitation term(s), just as usual bosonization methods\cite{4,5,6,7,8}, 
it can exactly
treat the system. b). For an interacting-free electron system, it is exact 
for dealing with the crossover excitation term(s). c). With this method,
the problem of the strongly correlated electron systems ends in to solve the
Eikonal-type equations, which can be exactly solved by a series expansion
of the boson fields $Q^{0}_{\sigma}(x,t)$, and/or by a computer calculations.
d). For some integrable systems, such as the quantum sine-Gordon model and
the Hubbard model at half-filling,
at one-loop approximation (i.e., only keeping the linear terms of the
Eikonal-type equations), it can give the correct results that are 
qualititively consistent with the exact solution of the systems for both
the weak and strong electron interactions.

It is the key point of the eigen-functional bosonization method to 
solve the Eikonal-type equations of the eigen-functional, and
use these eigen-functionals to calculate the electron Green's functional.
With the electron Green's functional, we can calculate the ground state energy
and the effective action of the system as well as correlation functions.
This method can also be used to 
treat one-dimensional electron-phonon interaction systems, 
two coupled spin-chain or quantum-wire systems, and one-dimensional disorder 
interacting electron systems. It can also be extended to two- and 
three-dimensional electron systems.

\end{document}